\documentclass[aps,twocolumn,showpacs]{revtex4}

\newcommand{\be}{\begin{equation}}
\newcommand{\ee}{  \end{equation}}
\newcommand{\ba}{\begin{eqnarray}}
\newcommand{\ea}{  \end{eqnarray}}

\begin{document}

\title{Universal Chaotic Scattering on Quantum Graphs}

\author{Z. Pluha\v r$^a$ and H. A. Weidenm{\"u}ller$^b$}
\email{Hans.Weidenmueller@mpi-hd.mpg.de}
\affiliation{$^a$Faculty of Mathematics and Physics, Charles University, 180 00 Praha 8, Czech Republic \\ $^b$Max-Planck-Institut f{\"u}r Kernphysik, 69029 Heidelberg, Germany}

\begin{abstract}We calculate the $S$--matrix correlation function for
  chaotic scattering on quantum graphs and show that it agrees with
  that of random--matrix theory (RMT). We also calculate all higher
  $S$--matrix correlation functions in the Ericson regime. These, too,
  agree with RMT results as far as the latter are known. We conjecture
  that our results give a universal description of chaotic scattering.

\end{abstract}


\maketitle

{\it Purpose.} Closed quantum systems that are chaotic in the
classical limit possess universal spectral fluctuation properties.
Depending on symmetry, these coincide with the fluctuation properties
of one of Dyson's three canonical random--matrix
ensembles~\cite{Dys62}. These statements, originally formulated in the
form of a conjecture~\cite{Boh84}, have since been demonstrated for
the two--point level correlator of general chaotic
systems~\cite{Heu04,Heu07} and of chaotic quantum graphs~\cite{Gnu04}.

For open chaotic quantum systems, the fluctuation properties of the
scattering matrix ($S$--matrix) are at issue, quantified in terms of
the totality of $S$--matrix correlation functions. As for closed
systems, it would be desirable to establish (at least) the complete
equivalence between the $S$-matrix two--point correlation function for
chaotic scattering with that of random--matrix theory (RMT) given in
Ref.~\cite{Ver85} for the orthogonal and in Ref.~\cite{Fyo05} for the
unitary case. However, we are not aware of any analytical results for
$S$--matrix correlations for chaotic scattering.

In the present Letter, we start to fill that gap. For the case of
orthogonal symmetry, we calculate the $S$--matrix two--point
correlation function analytically for open chaotic quantum graphs. We
show that our result coincides with that~\cite{Ver85} of RMT. We also
calculate all higher $S$--matrix correlation functions in the Ericson
regime. These coincide with RMT results as far as the latter are
known~\cite{Ver85, Aga75, Dav88}. In that regime, the $S$--matrix
elements are supposed~\cite{Eri60, Bri63} to have a Gaussian
distribution. Our results show that this is the case only for strong
absorption in all channels. We conjecture that our results are
universal, i.e., apply to chaotic scattering in general.

We focus attention on chaotic quantum graphs because here the
semiclassical expansion is exact and scattering theory is particularly
transparent. Chaotic scattering on quantum graphs was introduced in
Refs.~\cite{Kot00,Kot03} where many of its properties were displayed
with the help of numerical simulations, see also Refs.~\cite{Kot01}.

{\it Scattering Matrix.}  Our presentation is self--contained but
largely follows the developments of Ref.~\cite{Kot03}. A graph is a
system of $V$ vertices labelled $\alpha, \beta, \ldots$ that are
linked by $B$ bonds. For simplicity of notation we assume that every
vertex $\alpha$ is linked by a single bond ($\alpha \beta$) to every
other vertex $\beta$ (``completely connected graph''). Then the number
of bonds is $B = V (V - 1) / 2$.  Our results remain valid, however,
if some bonds are missing, see the discussion under ``massive modes''
below. The lengths $L_b$ of all bonds $b = (\alpha \beta)$ are assumed
to be similar (so that $L_{\min} \leq L_b \leq L_{\rm max}$ for all
$b$) and incommensurate.  That assumption is neccessary for the graph
to be chaotic. A number $\Lambda \geq 1$ of vertices is linked by a
single bond each (a ``lead'') to infinity. The number $\Lambda$ of
leads defines both, the number of scattering channels and the
dimension of the $S$--matrix. In analogy to the RMT approach (where
the dimension $N$ of the Hamiltonian matrix is taken to infinity while
the number of channels is kept fixed) we let $V \to \infty$ but keep
$\Lambda$ fixed.

On each bond or lead, waves propagate freely with wave number $k$ (the
same for all bonds/leads), and the wave function is a linear
combination of amplitudes $\exp \{ i k x \}$ and $\exp \{ - i k x \}$
where $x$ is the distance to one of the vertices attached to the
bond/lead. For all bonds/leads, the coefficients of the linear
combination are determined by boundary conditions specified in terms
of $V$ matrices $\Gamma^{(\alpha)}$ of dimension $V$ defined for each
vertex $\alpha$.  The matrix $\Gamma^{(\alpha)}$ expresses the
outgoing amplitudes on the lead and on each of the $V - 1$ bonds
attached to vertex $\alpha$ (written symbolically as ${\cal O}$) in
terms of the incoming amplitudes on the same or any other bond or the
lead attached to $\alpha$ (written symbolically as ${\cal I}$) so that
${\cal O} = \Gamma^{(\alpha)} {\cal I}$. With $\beta, \gamma \neq
\alpha$ the unitary and symmetric matrix $\Gamma^{(\alpha)}$ has the
form
\be
\Gamma^{(\alpha)} = \left(
  \matrix{ \rho^{(\alpha)} & \tau^{(\alpha)}_\gamma \cr
    \tau^{(\alpha)}_\beta & \sigma^{(\alpha)}_{\beta \gamma} \cr}
\right) \ .
\label{1}
\ee
Here $\rho^{(\alpha)}$ describes backscattering on lead $\alpha$,
$\tau^{(\alpha)}_\beta$ describes scattering from bond $(\alpha
\beta)$ to lead $\alpha$ or vice versa, and $\sigma^{(\alpha)}_{\beta
  \gamma}$ describes scattering from bond $(\alpha \beta)$ to bond
$(\alpha \gamma)$ or vice versa. In general, the matrix
$\sigma^{(\alpha)}$ is symmetric and subunitary. For a vertex $\alpha$
without lead the first row and column of $\Gamma^{(\alpha)}$ are
lacking, we have $\Gamma^{(\alpha)} = \sigma^{(\alpha)}$, and
$\sigma^{(\alpha)}$ is symmetric and unitary.

Given a single incident wave in channel $\alpha$ only, these boundary
conditions completely define the total wave function.  The amplitude
of the outgoing wave in channel $\beta$ is the element $S_{\alpha
  \beta}(k)$ of the symmetric and unitary scattering matrix.  To write
$S_{\alpha \beta}(k)$ explicitly, we define the symmetric subunitary
block--diagonal matrix $\Sigma$ of dimension $2 B = V (V - 1)$. Each
of the $V$ diagonal blocks of dimension $(V - 1)$ carries one of the
matrices $\sigma^{(\alpha)}$, $\alpha = 1, \ldots, V$. All other
elements of $\Sigma$ vanish. This defines the ``vertex
representation'' $\Sigma^{(V)}$ of $\Sigma$. The ``bond
representation'' $\Sigma^{(B)}$ is obtained by a reordering of rows
and columns. We assign to every bond $(\alpha \beta)$ the direction $d
= +$ ($d = -$) if $\alpha > \beta$ ($\alpha < \beta$, respectively).
We arrange the $B$ bonds with positive (negative) direction in
lexicographical order and label them consecutively with a running
index $b = 1, \ldots, B$. Then every directed bond is uniquely defined
by $(b, d)$. The total number of directed bonds is $2 B$. The map
$\sigma^{(\alpha)}_{\beta \gamma} \to \sigma_{(\alpha \beta), (\alpha
  \gamma)}$ defines the bond representation of $\sigma^{(\alpha)}$
and, thus, the bond representation $\Sigma^{(B)}$ of $\Sigma$. The map
$\tau^{(\alpha)}_\beta \to \tau_{(\alpha \beta)}$ similarly defines
the bond representation of the vector ${\cal T}$. In bond
representation we define the diagonal matrix $\exp \{ - i k {\cal L}
\}$ with elements $\delta_{b b'} \delta_{d d'} \exp \{ - i k L_b
\}$. Diagonal elements in locations that differ only in the sign of
$d$ are pairwise equal. Written somewhat symbolically the $S$--matrix
is
\be
S_{\alpha \beta}(k) = \delta_{\alpha \beta} \rho^{(\alpha)} + \big(
{\cal T} {\cal W}^{-1} {\cal T} \big)_{\alpha \beta}
\label{2}
\ee
where ${\cal W} = \exp \{ - i k {\cal L} \} - \Sigma^{(B)}$. Expanding
${\cal W}^{-1}$ in powers of $\Sigma^{(B)}$ we obtain a simple
physical interpretation of Eq.~(\ref{2}). The term containing the
$n$th power of $\Sigma^{(B)}$ is the sum of all semiclassical
trajectories that connect the vertices $\alpha$ and $\beta$ via
passage through $(n + 1)$ bonds. Each of the traversed bonds $(b d)$
yields the factor $\exp \{ i k L_{b} \}$.

{\it Averages.} The average over $k$ (indicated by angular brackets)
is taken over a $k$--interval that is larger than the minimum
difference between any two $L_b$'s. Because of the incommensurability
of the $L_{b}$, that average is equivalent~\cite{Gnu04} to a phase
average: For any function $F$ we have $\langle F[ \exp \{ i k L_{b}
\}] \rangle = (1 / (2 \pi)) \int_0^{2 \pi} {\rm d} \phi_{b} F [ \exp
\{ i \phi_{b} \} ]$. Then Eq.~(\ref{2}) implies $\langle S_{\alpha
  \beta} \rangle$ $= \delta_{\alpha \beta}
\rho^{(\alpha)}$~\cite{Kot03} and can, thus, be read as $S = \langle S
\rangle + S^{\rm fl}$ where the fluctuating part is $S^{\rm fl} =
{\cal T} {\cal W}^{-1} {\cal T}$. It also follows that the average of
the product of any number of $S$--matrix elements is equal to the
product of the averages (as in RMT).

{\it Supersymmetry and saddle--point approximation.} The $S$--matrix
correlation function $(P, Q)$ is defined as the average of a product
of $P$ elements of $S^{\rm fl}$ with arguments $k + \kappa_p$, $p = 1,
\ldots, P$ and $Q$ elements of $S^{{\rm fl} *}$ with arguments $k -
\tilde{\kappa}_q$, $q = 1, \ldots, Q$. Without loss of generality we
assume $P \geq Q \geq 1$. Since $S^{\rm fl} = {\cal T} {\cal W}^{-1}
{\cal T}$, it suffices to work out
\be
\bigg\langle \prod_{p = 1}^P {\cal W}^{-1}_{b_p d_p, b'_p d'_p}(k +
\kappa_p) \prod_{q = 1}^Q \big({\cal W}^{-1}_{b_q d_q, b'_q d'_q}(k
- \tilde{\kappa}_q)\big)^* \bigg\rangle \ .
\label{4}
\ee
We generalize the approach of Refs.~\cite{Gnu04, Gnu08}. Using
supersymmetry~\cite{Efe83, Ver85}, the correlator~(\ref{4}) is written
as the $(P + Q)$--fold derivative of the average of a generating
function $G$ (a superintegral). The average over $k$ is calculated as
a phase average over all $\phi_{b}$ with the help of the
colour--flavour transformation~\cite{Zir96} in its most general form
(for $P \neq Q$). Integrating out the original integration variables
gives
\be
\langle G \rangle = \int {\rm d} (\tilde{Z}, Z) \exp \{ - {\cal
A}(\tilde{Z}, Z) \}
\label{5}
\ee
where ${\cal A}(\tilde Z, Z)$ is the action
\ba 
&& {\cal A}(\tilde Z, Z) = - {\rm STr} \ln ( 1 - Z \tilde{Z}) +
\frac{1}{2} {\rm STr} \ln (1 - Z z Z^{\tau} z) \nonumber \\
&& \qquad + \frac{1}{2} {\rm STr} \ln (1 - {\cal B}_{+}^{-1} \tilde
Z ^\tau z\, ({\cal B}_{-}^\dag )^{-1} \tilde Z z) \ .
\label{6}
\ea
Here ${\rm STr}$ is the supertrace. All matrices are defined in bond
and in retarded--advanced representation. In the retarded (advanced)
sector, the matrix dimension is $8 B P$ ($8 B Q$, respectively), a
factor $4$ arising from supersymmetry. The matrix $Z$ ($\tilde{Z}$)
fills the upper (lower) non--diagonal block of the retarded--advanced
sector, respectively, and is rectangular for $P > Q$. Both $Z$ and
$\tilde{Z}$ are diagonal in bond space. The matrices $Z$ and $Z^\tau$
are related as in Ref.~\cite{Gnu04}. The matrix $z$ is diagonal with
diagonal elements given by $\exp \{ i \kappa_p L_b/2 \}$, $p = 1,
\ldots, P$ in the retarded sector and by $\exp \{ i \tilde{\kappa}_q
L_b/2 \}$, $q = 1, \ldots, Q$ in the advanced sector. The matrix
${\cal B}_+$ (${\cal B}_-$) is only defined for the retarded (the
advanced) sector, respectively. ${\cal B}_+^{-1}$ is block--diagonal
with regard to the index $p = 1, \ldots, P$ and in each block given by
$\Sigma^{(B)}_{b d, b' d'} + \sigma^s_3 A^{(p)}$, and correspondingly
for ${\cal B}^{-1}_-$. Here $\sigma^s_3$ is the third Pauli matrix in
superspace, and $A^{(p)}$ denotes the source term needed to generate
by differentiation of $G$ the matrix element ${\cal W}^{-1}_{b_p d_p,
  b'_p d'_p}$. The integration measure in Eq.~(\ref{5}) is the flat
Berezinian. Up to this point our results are exact.

We calculate $\langle G \rangle$ using the saddle--point
approximation, putting $z = 1$ and $A^{(j)} = 0$ for all $j$. As in
Ref.~\cite{Gnu04}, variation of the resulting action with respect to
$Z$ and $\tilde{Z}$ yields the saddle--point equation $(1 - Z
\tilde{Z})^{-1} Z = (1 - \Sigma_+ Z \Sigma^{*}_- \tilde{Z})^{-1}
\Sigma_+ Z \Sigma^{*}_-$. Here $\Sigma_+$ is block--diagonal and in
each block labelled $p = 1, \ldots, P$ given by $\Sigma^{(B)}$, and
correspondingly for $\Sigma_-$. The saddle--point equation holds if $Z
\Sigma^*_- = \Sigma^*_+ Z$ and if $\Sigma^{(B)} \Sigma^{(B) *} =
1$. We write $\Sigma^{(V)} \Sigma^{(V) *} = 1 + \delta^{(V)}$,
$\Sigma^{(B)} \Sigma^{(B) *} = 1 + \delta^{(B)}$. As done in RMT, we
first suppress $\delta^{(V)}$ and $\delta^{(B)}$ (both of which are
caused by coupling to the channels). We then work out the ensuing
corrections to the saddle--point solution exactly in Eq.~(\ref{10})
below. To satisfy $Z \Sigma^*_- = \Sigma^*_+ Z$ we follow
Ref.~\cite{Gnu04} and write the universal saddle--point solution $Y$
as $\delta_{b b'} \delta_{d d'} Y_{p t s, q t' s'}$, and
correspondingly for $\tilde{Y}$.

Corrections to the saddle--point action are due to deviations from $z
= 1$, and from $\Sigma^{(B)} \Sigma^{(B) *} = 1$. Concerning the
former, we expand~\cite{Gnu04} $z$ and the action ${\cal A}$ around
the saddle--point value up to first order in $\kappa_p$ and
$\tilde{\kappa}_{q}$. With $\langle d_{\rm R} \rangle = (1 / \pi)
\sum_b L_b$ the average level density~\cite{Gnu04}, we obtain in the
exponent of Eq.~(\ref{5}) the ``symmetry--breaking term''
\be
SY = i \pi \langle d_{\rm R} \rangle \bigg( {\rm STr}_{p s t} \kappa
\frac{1}{1 - Y \tilde{Y}} + {\rm STr}_{q s t} \tilde{\kappa}
\frac{1}{1 - \tilde{Y} Y} \bigg) \ .
\label{8}
\ee
With $s, t$ labelling the superindices, the trace is only over the
subspaces indicated, the matrix $\kappa$ is $\delta_{s s'} \delta_{t
  t'} \delta_{p p'} \kappa_p$, and correspondingly for
$\tilde{\kappa}$. Deviations from $\Sigma^{(B)} \Sigma^{(B) *} = 1$
are calculated by putting $z = 1$ and dropping the source terms. We
use Eq.~(\ref{1}), suppress the index $\alpha$, and take $\rho$ to be
real (as in RMT, that suppresses all elastic scattering phase
shifts). Since $\Gamma$ is unitary and symmetric it can be unitarily
transformed into
\be
\left( \matrix{ \rho & \exp \{ - i \phi_1 \} T^{1/2} & 0 \cr
\exp \{ - i \phi_1 \} T^{1/2} & - \rho \exp \{ - 2 i \phi_1 \} & 0 \cr
0 & 0 & \delta_{\mu \nu} \exp \{ i \phi_\mu \} \cr} \right) \ . 
\label{9}
\ee
Here $T = 1 - \rho^2$ is the transmission coefficient, the indices
$\mu, \nu$ run from $2$ to $V - 1$, and the phases $\phi_1, \phi_\mu$
are real and arbitrary.  Eq.~(\ref{9}) shows that $\sigma \sigma^*$
differs from the unit matrix only in the first diagonal element which
is $1 - T$. Using that for all channels $\alpha$ we obtain in the
exponent of Eq.~(\ref{5}) the ``channel--coupling term''
\be
CH = - \frac{1}{2} \sum_{\alpha = 1}^V {\rm STr}_{p s t} \ln \bigg(
1 + T^{(\alpha)} \frac{Y \tilde{Y}}{1 - Y \tilde{Y}} \bigg) \ .
\label{10}
\ee
Actually the sum extends only over the $\Lambda \ll V$ vertices
coupled to a lead. Collecting everything we find
\be
\langle G \rangle = \int {\rm d} (Y, \tilde{Y}) \bigg( ... \bigg)
\exp \{ SY + CH \} \ .
\label{11}
\ee
The term in big round brackets contains the source terms.

{\it Massive Modes}. In Eq.~(\ref{11}) we have neglected the massive
modes because our interest in the present paper is focused on
generating universal results for chaotic scattering without using the
framework of RMT. Massive modes were investigated in
Refs.~\cite{Gnu04, Gnu08} for closed graphs. For the statistics of
eigenfunctions it was found~\cite{Gnu08} that sequences of graphs with
monotonically increasing $V$ are quantum ergodic (i.e., the massive
modes do not contribute) if the spectrum of eigenvalues of the
analogue of our matrix $|\Sigma^{(B)}_{b d, b' d'}|^2$ asymptotically
($V \to \infty$) possesses a gap separating it from
zero. Wave--function statistics is known~\cite{Mit10} to be important
for $S$--matrix fluctuations.  Therefore, we conjecture that for
$\Lambda \ll V$ ($\Lambda$ fixed and $V \to \infty$) that criterion
applies in the present case to the spectrum of our $|\Sigma^{(B)}_{b
  d, b' d'}|^2$. A proof would require a detailed investigation.

{\it Two--point Function}. For the correlation function $(1, 1)$ the
matrices $Y$ and $\tilde{Y}$ are both square matrices of dimension
four, and it is straightforward to work out the source terms in
Eq.~(\ref{11}). Lack of space does not permit us to present any
details. Suffice it to say that using the transformations $t_{1 2} = Y
(1 - Y \tilde{Y})^{- 1/2}$, $t_{2 1} = \tilde{Y} (1 - Y \tilde{Y})^{-
  1/2}$, writing $\langle S_{\alpha \alpha} \rangle$ for
$\rho^{(\alpha)}$ and $T^{(\alpha)} = 1 - | \langle S_{\alpha \alpha}
\rangle|^2$, and replacing the wave--number arguments of $S$ by
energies, the resulting expressions for $\langle S^{\rm fl}_{\alpha
  \beta}(k + \kappa) S^{{\rm fl} *}_{\gamma \delta}(k - \kappa)
\rangle$ become formally identical to the corresponding terms in
Eq.~(7.23) of Ref.~\cite{Ver85} for all values of the number $\Lambda
= 1, \ldots, V$ of channels. For the terms $SY$ and $CH$ in
Eqs.~(\ref{8}) and (\ref{10}) that can be checked directly. In the
source terms, the phase $\phi_1$ cancels out. These facts establish
the equivalence of the two--point functions of RMT and of chaotic
scattering on quantum graphs.

{\it Ericson Regime.} That regime is defined by the condition
$\sum_\alpha T^{(\alpha)} \gg 1$. The cross section for chaotic
scattering is expected to display Ericson
fluctuations~\cite{Eri60,Bri63}. Numerical simulations~\cite{Kot03}
have confirmed that expectation. Eq.~(\ref{11}) allows us to determine
the leading terms in an asymptotic expansion in inverse powers of
$\sum_\alpha T^{(\alpha)}$ of all $(P, Q)$--correlation functions and,
thus, the complete distribution of $S$--matrix elements in the Ericson
regime. The asymptotic terms are obtained~\cite{Ver86} by keeping in
Eq.~(\ref{11}) only terms of lowest order in $Y$, $\tilde{Y}$. For $SY
+ CH$ we obtain
\be
- \frac{1}{2} \sum_{p q} \bigg( \sum_\alpha T^{(\alpha)} - 2 i \pi
\langle d_{\rm R} \rangle (\kappa_p + \tilde{\kappa}_{q}) \bigg)
{\rm STr}_{s t} (Y_{p q} \tilde{Y}_{q p}) \ . 
\label{13}
\ee
For the two--point function, the calculation~\cite{Ver86} yields
\be
\langle S^{\rm fl}_{\alpha \beta}(k + \kappa) S^{{\rm fl} *}_{\gamma
\delta}(k - \tilde{\kappa}) \rangle = \frac{(\delta_{\alpha \gamma}
\delta_{\beta \delta} + \delta_{\alpha \delta} \delta_{\beta \gamma}
) T^{(\alpha)} T^{(\gamma)}} {\sum_\tau T^{(\tau)} - 2 i \pi
\langle d_{\rm R} \rangle (\kappa + \tilde{\kappa})} \ . 
\label{14}
\ee
With $T^{(\alpha)} = 1 - |\langle S_{\alpha \alpha} \rangle|^2$ and
the replacement of wave numbers by energies, this is exactly the
expression obtained for RMT in Refs.~\cite{Aga75, Wei84}. For the
general $(P, Q)$ correlation function we need to find the
leading--order contribution to the source terms. We expand the last
term in Eq.~(\ref{6}) with $\tilde{Z}^\tau \to Y$ and $\tilde{Z} \to
\tilde{Y}$, retaining only terms linear in $Y$ and $\tilde{Y}$. (Only
these are of the form $\sum_{p q} Y_{p q} \tilde{Y}_{q p}$ which,
according to Eq.~(\ref{13}), gives the leading--order contribution).
We need $P$ source terms from ${\cal B}_+$. Expanding the exponential
we keep the term $(1 / (2^{P} P!))  [ \sum_{p q} {\rm STr}_{s t} \big(
({\cal B}^{-1}_+)_p Y_{p q} ({\cal B}^{\dag -1}_+)_{q} \tilde{Y}_{q p}
\big) ]^P$. No two source terms in ${\cal B}^{-1}_+$ may have the same
labels. That gives $(1 / 2^P) \prod_{p} [ \sum_{q} {\rm STr}_{s t}
\big( ( \sigma^s_3 A^{(p)} Y_{p q} ({\cal B}^{\dag -1}_+)_{q}
\tilde{Y}_{q p} \big)]$. Since all source terms in ${\cal B}^{\dag
  -1}_-$ must also be different, the sum over $q$ goes for $P = Q$
over all permutations of $q = 1, \ldots, Q$. Equivalently we may keep
all $q$ fixed and sum over all permutations of $p = 1, \ldots, P$. For
$P > Q$ there are $P - Q$ source terms $A^{(p)}$ in ${\cal B}^{\dag
  -1}_+$ that do not have a counterpart in ${\cal B}^{\dag -1}_-$. For
these ${\cal B}^{\dag -1}_-$ is replaced by $\Sigma$. Each of the
resulting supertraces corresponds to one of the factors in the flat
integration measure ${\rm d}(Y, \tilde{Y}) = \prod_{p q} {\rm d}(Y_{p
  q}, \tilde{Y}_{q p})$. Therefore and because of Eq.~(\ref{13}), each
superintegral factorizes into $P Q$ terms, each factor characterized
by the pair $(p, q)$ of indices. In factors that do not carry any
source terms the superintegration gives unity. The integration over
those supertraces which carry both factors $A^{(p)}$ and $A^{(q)}$
yields the asymptotic form~(\ref{14}) of the average of a pair of
$S$--matrix elements. For the $P - Q$ unpaired source terms $A^{(p)}$
the superintegration gives a non--vanishing contribution for $\alpha_p
= \beta_p$ only. The resulting factor is
\be
{\cal F}_{\alpha_p}(\kappa_p) = - \sum_{q = 1}^Q \frac{T^{(\alpha_p)}
\langle S_{\alpha_p \alpha_p} \rangle}{\sum_\gamma T^{(\gamma)} - 2
i \pi (\kappa_{p} + \tilde{\kappa}_{q}) \langle d_{\rm R} \rangle}
\label{15}
\ee
where again the phase $\phi_1$ in Eq.~(\ref{9}) cancels out. The sum
over $q$ arises because in the advanced block, the matrix ${\cal
  B}^{\dag -1}_-$ carries the same entry $\Sigma$ in every block
labelled $q$. We suppress all arguments $k$ for brevity. That gives
\ba
&& \bigg\langle \prod_{p = 1}^P S^{\rm fl}_{\alpha_p
\beta_p}(\kappa_p) \prod_{q = 1}^Q S^{{\rm fl} *}_{\alpha'_{q}
\beta'_{q}}(- \tilde{\kappa}_{q}) \bigg\rangle =  \sum_{\rm sel}
\prod_{j = 1}^{P - Q} {\cal F}_{\alpha_{p_j}}(\kappa_{p_j})
\nonumber \\
&& \times \sum_{\rm perm} \prod_{q = 1}^Q \bigg\langle S^{\rm
fl}_{\overline{\alpha}_q \overline{\beta}_q} (\overline{\kappa}_q)
S^{{\rm fl} *}_{\alpha'_{q} \beta'_{q}}(- \tilde{\kappa}_{q})
\bigg\rangle \ .
\label{16}
\ea
The sum with index ``sel'' goes over all ${P \choose P - Q}$
possibilities to select $(P - Q)$ matrix elements $S^{\rm fl}$ from
the first factor on the left--hand side. These give rise to the first
product which vanishes unless all selected elements are diagonal. The
remaining $Q$ elements $S^{\rm fl}$, symbolically written as $S^{\rm
  fl}_{\overline{\alpha}_j \overline{\beta}_j}(\overline{\kappa}_j)$
with $j = 1, \ldots, Q$, appear as first factors in the angular
brackets on the right--hand side. The sum with index ``${\rm perm}$''
extends over all permutations of these elements. Each of the terms in
angular brackets on the right--hand side is equal to the asymptotic
form~(\ref{14}) of the two--point function.

Eq.~(\ref{16}) gives the asymptotic form of all $S$--matrix
correlation functions $(P, Q)$ and, thus, the complete distribution of
the $k$--dependent scattering matrix in the Ericson regime. For the
$(2, 2)$ correlation function it agrees with the result of
Ref.~\cite{Aga75}. If $\langle S_{\alpha \alpha} \rangle = 0$ for all
$\alpha$ (strong absorption in all channels), the factors ${\cal F}$
all vanish, the correlation functions $(P, Q)$ vanish for $P \neq Q$,
and for $P = Q$ have the form characteristic of a Gaussian random
process. In particular, all elements of $S$ have a Gaussian
distribution centered at zero, and cross--section fluctuations have
the form predicted in Refs.~\cite{Eri60,Bri63}. If $\langle S_{\alpha
  \alpha} \rangle \neq 0$ in some channel $\alpha$, that fact and the
unitarity constraint $|S_{\alpha \alpha}| \leq 1$ distort the Gaussian
distribution of $S_{\alpha \alpha}$. This is the cause of the
occurrence of the factor ${\cal F}_\alpha$ (Eq.~(\ref{15})). The
factor shows that the distortion is biggest for $|\langle S_{\alpha
  \alpha} \rangle| = 1 / \sqrt{2}$. For cross--section correlation
functions, interest is focussed on the (2,1) correlation function. It
was first noted in Refs.~\cite{Dav88} that this function differs from
zero (the result given there agrees aympotically with our
Eq.~(\ref{16})).  Implications of that fact for cross--section
fluctuations have been discussed in Ref.~\cite{Die10}.

{\it Conclusions.} For chaotic scattering on quantum graphs we have
derived formal analytical expressions for all $(P, Q)$ correlation
functions of the $S$--matrix. These were used to show that the $(1,
1)$ correlation function is identical to the one obtained from RMT,
and to calculate for all $(P, Q)$ explicit expressions in the Ericson
regime. The latter agree with RMT results as far as these are known
and yield the complete $S$--matrix distribution function in that
regime. (It may perhaps be possible to obtain from Eq.~(\ref{11})
explicit expressions also for $(2, 1)$ and $(2, 2)$.) We conjecture
that our results apply asymptotically for sequences of quantum graphs
that when closed are quantum ergodic.

Two facts suggest that our results are universal, i.e., hold for
quantum--chaotic scattering in general: The agreement of our results
with those of RMT, and the agreement of the two--point function for
closed graphs~\cite{Gnu04} with that of general closed chaotic
systems~\cite{Heu07}. Therefore, we conjecture that quantum--chaotic
scattering and the RMT approach to scattering are completely
equivalent.

ZP acknowledges support by the Czech Ministry of Education under
Project MSM 0021620859. The authors are grateful for valuable comments
to A. Altland, S. Gnutzmann, J. Kvasil, P. Cejnar, and U. Smilansky.


\end{document}